\documentclass[onecolumn]{article}
\usepackage[utf8]{inputenc}
\usepackage{graphicx}
\usepackage{color}
\usepackage{amsmath}
\usepackage{subfig}
\usepackage{float}
\usepackage{multirow}
\usepackage{flafter}
\usepackage{authblk}

\title{Effects of Pb doping on structural and electronics properties of Bi$_2$Sr$_2$Ca$_2$Cu$_3$O$_{10}$ }
\author{J.A. Camargo-Martínez}

\affil{Grupo de Investigación en Ciencias Básicas, Aplicación e Innovación - CIBAIN\\ 
Fundación Universitaria Internacional del Trópico Americano - Unitrópico, Yopal - Colombia}

\author{R. Baquero}
\affil{Departamento de Física, CINVESTAV-IPN, Av. IPN 2508, 07360 México}

\begin{document}

\maketitle

\begin{abstract} 

Pb doping effect in the Bi$_2$Sr$_2$Ca$_2$Cu$_3$O$_{10}$ compound (Bi2223) on the structural and electronic properties were investigated, using the Local Density (LDA) and Virtual Crystal (VCA) approximations within the framework of the Density Functional Theory (DFT), taking as reference the procedure implemented by H.Lin {\it et al.} in the Bi2212 compound [{\it Phys. Rev. Lett.} {\bf 96} (2006) 097001]. Results show that, the incorporation of Pb-dopant in Bi2223 lead a rigid displacement of the Bi/Pb-O bands toward higher energies, with a null contribution at the Fermi level, around the high symmetry point $\overline{\text{M}}$ in the irreducible Brillouin zone, for Pb doping concentration equal to or more than 26\%, avoiding the presence of the so-called Bi-O {\it pockets} in the Fermi surface, in good agreement with angle-resolved photoemission spectroscopy (ARPES) and nuclear magnetic resonance (NMR) experiments, although a slight metallic character of the Bi-O bonds is still observed which would disagree with some experimental reports. The calculations show that the changes on the structural properties are associated to the presence or absence of the Bi-O {\it pockets} in the Fermi surface.\\

\end{abstract}




\section{Introduction}

The Bi$_2$Sr$_2$Ca$_2$Cu$_3$O$_{10}$ compound (Bi2223) is a high-temperature superconductor of the cuprate bismuth family with general formula Bi$_2$Sr$_2$Ca$_{n-1}$Cu$_n$O$_{10}$, which differ by the number of CuO$_2$ planes per unit cell. Resistivity, susceptibility and magnetization experimental measurements in the Bi2223 compound ($n=$ 3) show a transition to the superconducting state at $\sim110$ K~\cite{A1,A2}. In despite of the fact that this compound has a very high viability for its use in industrial and technological applications~\cite{A,B,C}, we found surprisingly few theoretical reports.

In a previous work~\cite{3a} we calculated the electronic properties of Bi2223 with tetragonal phase (I4/mmm), where the presence of electronic states associated with the Bi-O planes at the Fermi level ($E_F$) was evident. These are observed in the Fermi surface (FS) as small closed surfaces around the anti-nodal point $\overline{\text{M}}$ in the irreducible first Brillouin zone (IZB), in disagreement with experimental reports~\cite{5a}. This issue is known as the Bi-O {\it pockets} problem~\cite{7a,8a} and has been present in the literature since long ago~\cite{1a,1aA,2a,4a,6a,6aA}.

We have shown in a later work~\cite{3ax}, that small changes in the position of the oxygen atom associated with the Sr plane in Bi2223, induce a quasi-rigid displacement of the Bi-O bands towards higher energies avoiding its contribution at $E_F$ offering a solution to the Bi {\it pockets} problem in theoretical calculations. It is necessary to validate experimentally these atomic positions.

Previously, H.Lin {\it et al.} showed that in the Bi$_2$Sr$_2$Ca$_1$Cu$_1$O$_{8}$ compound (Bi2221) the Bi-O bands are lifted above $E_F$ when Pb-dopand is incorporated in concentrations greater than 22\%, avoiding the Bi-O {\it pockets}~\cite{7a,8a}. In that paper the authors state that this procedure has the same effect on the electronic properties of Bi2201 and Bi2223 compounds but without proof. 

In this paper, we present a study of Pb doping effects on the structural and electronic properties of (Bi$_{1-x}$Pb$_x$)$_2$Sr$_2$Ca$_2$Cu$_3$O$_{10}$ for $ 0 \leq x \geq 0.35$, using the ab initio virtual crystal approximation. We found that the Bi-O {\it pockets} disappear from the FS at a specific Pb doping, although a slight metallic character of the Bi-O bonds remains which would disagree with some experimental observations as we mentioned before.

\section{Method of Calculation}

The calculations were done using the full-potential linearized augmented plane wave method plus local orbital (FLAPW+lo)~\cite{10a} within the local density approximation (LDA), using the wien2k code~\cite{11a}. We used a plane-wave cutoff at $R_{mt}K_{max}=$ 8.0 and for the wave function expansion inside the atomic spheres, a maximum value for the angular momentum of $l_{max}=12$ with $G_{max}= 20$. We choose a $14\times14\times14$ k-space grid which contains 288 points within the IBZ. The muffin-tin sphere radii $R_{mt}$ (in atomic units) were chosen to be 2.3 for Bi, 2.0 for Sr, 1.9 for both Ca and Cu, and 1.5 for O.

Pb substitution by Bi atoms was considered within the framework of the virtual crystal approximation (VCA), where the Bi nuclear charge Z is replaced by the average of the Bi and Pb charge of what may be thought as an ``effective'' Bi/Pb atom~\cite{7a}. For this system it was found that the effective disorder parameter for the Bi-O states is $\Delta/W \sim$ 0.33, where $\Delta$ is the splitting of the Bi-O and Pb-O bands in Bi2223 and Pb2223, respectively and $W$ the bandwidth, thus ensuring that the system is far from being in the split-band limit~\cite{7a,8a,n21}, confirming that the VCA is a good approximation in this case.

\section{Results and discussion}

\subsection{Structural properties}

We studied the Pb doping effects on the structural properties of Bi2223 with a body-centered tetragonal structure (bct) and space group I4/mmm ($D_{4h}^{17}$). The structure consists of three Cu-O planes, one Cu1-O1 plane between two Cu2-O2 planes, with Ca atoms between them. Each Cu2-O2 plane is followed by a Sr-O3 and Bi-O4 planes in that order.

\begin{table}[h!]
\centering
\caption{Optimized lattice parameters and atomic coordinates relaxed for (Bi$_{1-x}$Pb$_x$)$_2$Sr$_2$Ca$_2$Cu$_3$O$_{10}$ with body-centered tetragonal structure and space group I4/mmm, for Pb concentrations of $x = $ 0.0, 0.2, 0.25, 0.26, 0.3 and 0.35. The experimental values were taken from ref.~\cite{G}.}
\begin{tabular*}{1.03\textwidth}{cccccccc}\hline\hline
&\multicolumn{7}{c}{Lattice parameters in \AA} \\
\cline{2-8}
Parameter     & Expt.        &0.0& 0.20   &  0.25 & 0.26    &   0.30 & 0.35  \\\hline
{\em a}	      &3.823(9)      &3.8060 &3.8058  &3.8078 & 3.8080  & 3.8084&3.80787\\
{\em c}       &37.074(5)     &37.407 &37.409  &37.372 & 37.369  &37.358 &37.3692\\\hline
{\em c/a}     &9.6976(2)     &9.8284 &9.8294  &9.8146 & 9.8134  &9.8094 &9.8137\\\hline\hline

&\multicolumn{7}{c}{Atomic coordinates} \\								
\cline{2-8}
Atom            & Expt.     &0.0 & 0.20  &  0.25 & 0.26   &  0.30 & 0.35  \\\hline
Bi$_{(1-x)}$Pb$_x$ &0,2109(6)&0,2018&0,2034&0,2033&0.2035 &0,2035&0,2031\\
Sr               &0,3557(2)  &0,3709&0,3693&0,3694&0.3685 &0,3687&0,3689\\
Ca               &0,4553(8)  &0,4581&0,4576&0,4576&0.4572 &0,4574&0,4575\\
Cu1              &0,0000     &0,0000&0,0000&0,0000&0.0000 &0,0000&0,0000\\
Cu2              &0,0976(4)  &0,0823&0,0833&0,0832&0.0840 &0,0837&0,0836\\
O1               &0,0000     &0,0000&0,0000&0,0000&0.0000 &0,0000&0,0000\\
O2               &0,0964(2)  &0,0828&0,0834&0,0832&0.0835 &0,0831&0,0830\\
O3               &0,1454(4)  &0,1477&0,1489&0,1488&0.1490 &0,1486&0,1482\\
O4               &0,2890(2)  &0,2996&0,2992&0,2995&0.2992 &0,2998&0,3002\\
\hline\hline
\end{tabular*}
\label{T1}
\end{table}

Starting from the experimental parameters taken from reference~\cite{G}, the $c/a$ ratio was optimized by minimizing the total energy
at a constant volume. Also the atomic coordinates were relaxed by minimizing the total force for each Pb concentration.

In Table~\ref{T1} the values of the lattice parameters and the atomic coordinates obtained in this work are reported. In all cases, the $c$ and $a$ lattice parameters calculated show differences below 1\% with respect to the reported experimental values. As a general behaviour it is observed that the calculated values of $c$ are always overestimated while those of $a$ are underestimated as compared to the experimental data.

\begin{figure}[htb!]
\begin{center}
\includegraphics[width=1\textwidth]{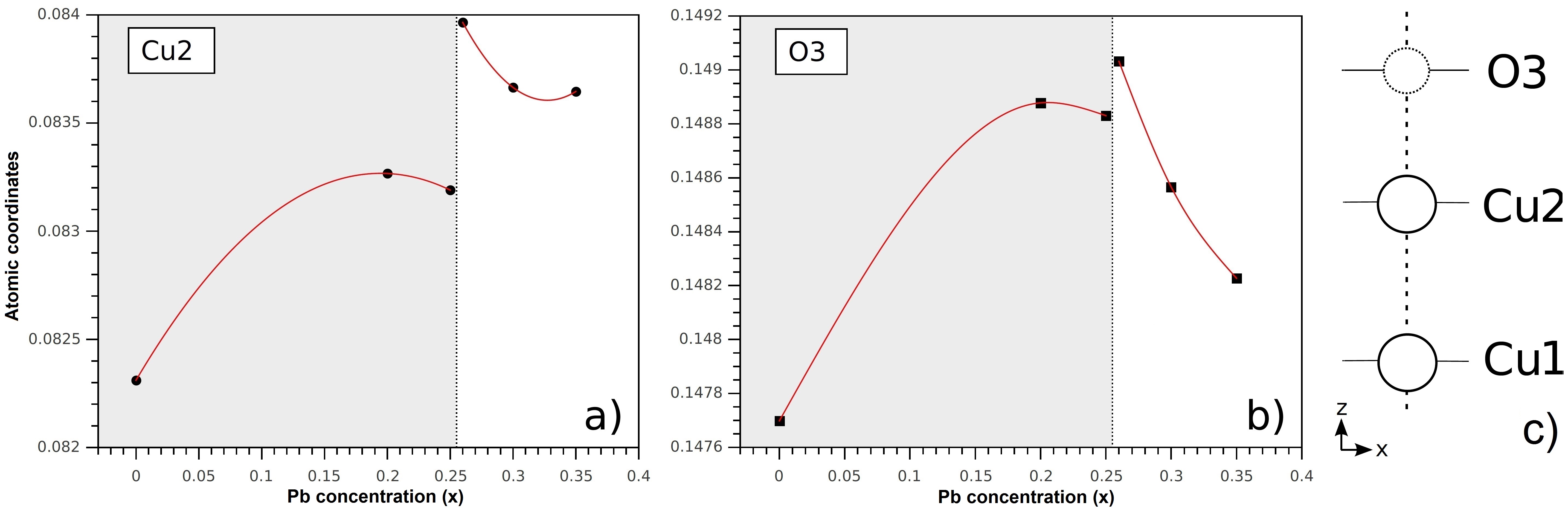}
\caption{(Color online) Pb doping effects on the atomic coordinates of a) Cu2 and b) O3 in the Bi$_2$Sr$_2$Ca$_2$Cu$_3$O$_{10}$ compound. c) Schematic positions of Cu1, Cu2 and O3 in the structure.}
\label{F1}   
\end{center}
\end{figure}

When the atomic coordinates are relaxed as a function of Pb concentration (see table~\ref{T1}), a very sharp discontinuity in the value of the internal coordinates appears around $x\sim$ 0255 dividing therefore the pb doping effect on the structure into two distinct regions as it can be observed in Fig.~\ref{F1} for the atoms Cu2 and O3.

It is found that the greater discontinuity occurs in the Cu2, which is interpreted as an important decrease in the internal parameter of Cu1 with Pb concentration (see Fig.~\ref{F1}a). O3 has a similar behaviour, its proximity to Cu2 is enhanced for Pb concentrations greater than $x=$ 0.25. The decrease in the Cu2-O3 distance has important effects on the electronic properties of Bi2223~\cite{3ax}.

\begin{figure}[!ht]
\begin{center}
\includegraphics[width=0.5\textwidth]{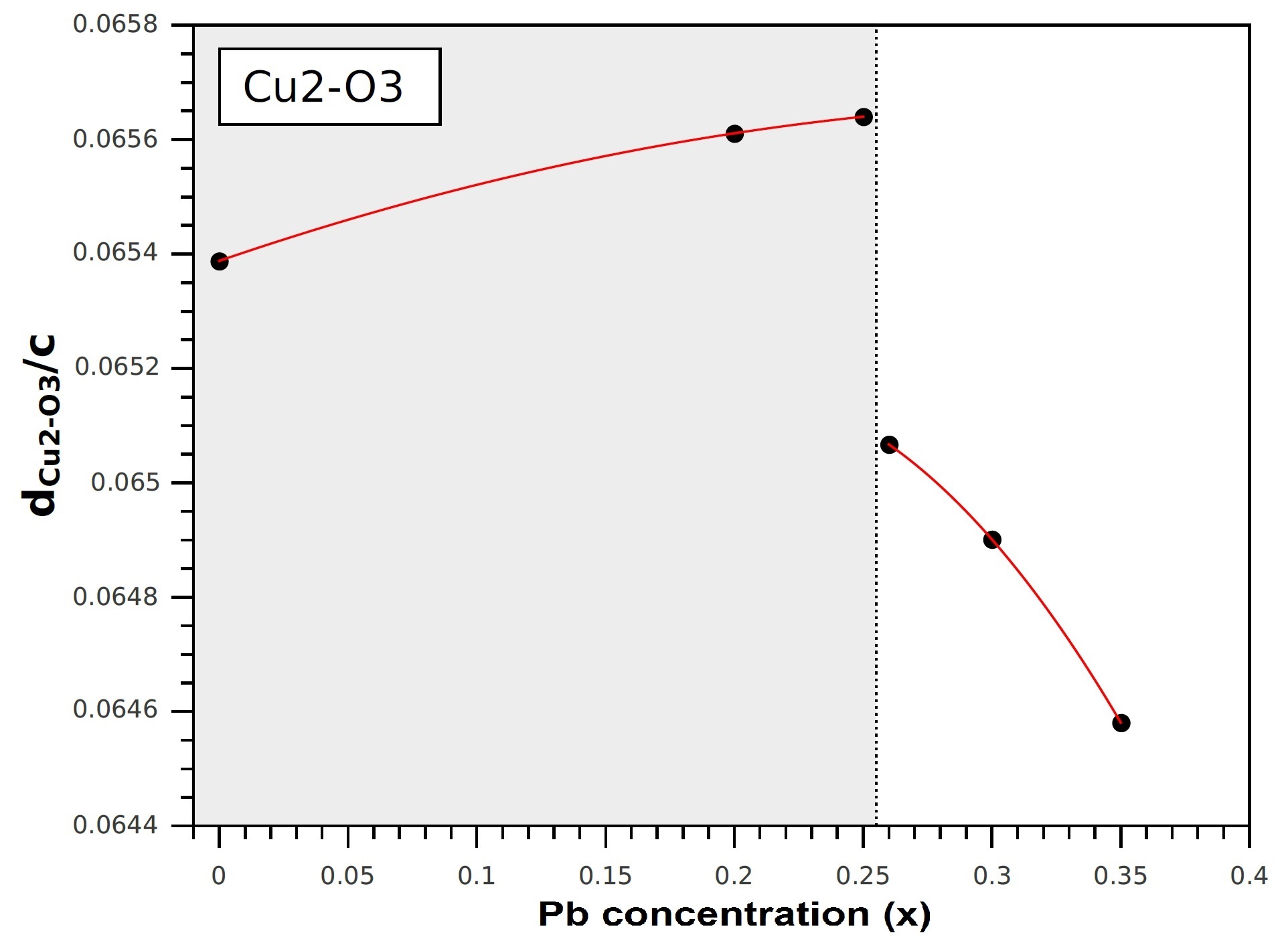}
\caption{(Color online) Pb doping effects on the Cu2-O3 distance (d$_{Cu2-O3}/c$) in the Bi$_2$Sr$_2$Ca$_2$Cu$_3$O$_{10}$ compound.}
\label{F2} 
\end{center}
\end{figure}

We show in Fig.~\ref{F2} the effects of Pb doping on the Cu2-O3 relative distance (d$_{Cu2-O3}/c$). Initially, doping induces a progressive increase of the Cu2-O3 distance.This behaviour changes completely for Pb concentrations greater than 0.25. For a Pb concentration of 0.26, i.e., Pb doping of 26\%, the Cu2-O3 distance is 2.43 \AA, which is 0.6 to 0.8 \AA ~ greater than the experimental values reported~\cite{G,dis1,dis2,dis3,dis4}. The difference with experiment is clear.

In Figs.~\ref{F1} and ~\ref{F2} two regions are distinguished, the gray region represent the presence of Bi-O {\it pockets} in the FS and the white area indicates the absence of them. In these regions the behaviour of the structural properties clearly shows notable differences associated to the presence (or absence) of the Bi-O {\it pockets}.

\subsection{Electronic properties}

\begin{figure}[!ht]
\begin{center}
\includegraphics[width=1\textwidth]{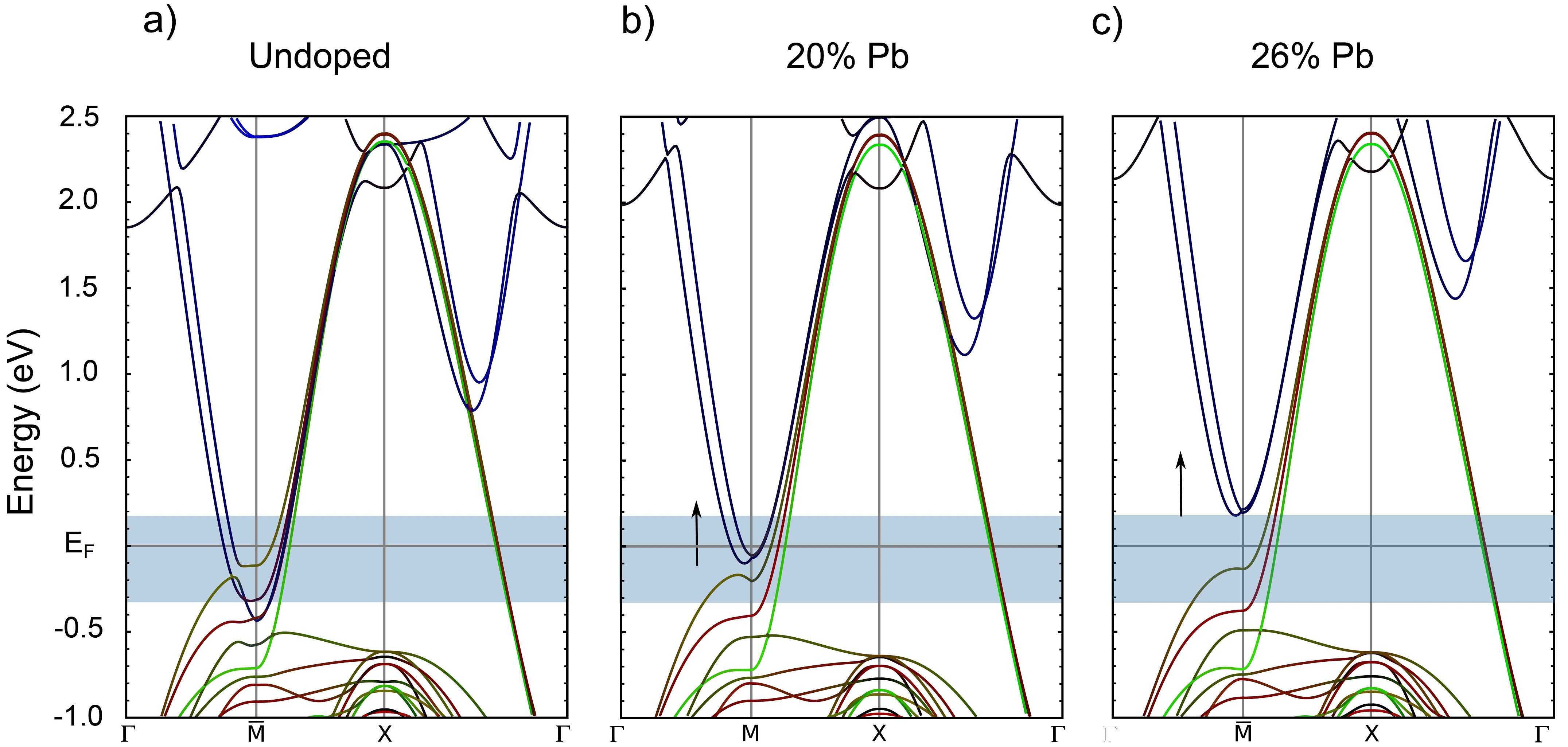}
\caption{(Color online) Pb doping effects on the band structure of the Bi$_2$Sr$_2$Ca$_2$Cu$_3$O$_{10}$ compound for the Pb concentration of a) 0.0 (Undoped), b) 0.2 (20\% Pb) and c) 0.26 (26\% Pb). The blue, red and green lines represent the Bi $p$, Cu2 $d$ and Cu1 $d$ states respectively. The hybridized states from these bands are represented by their respective color mixture and the black line represents the other states.}
\label{B} 
\end{center}
\end{figure}

The electronic properties of Pb-free and Pb-doped Bi2223 are different in three regions. In the first, Pb-free Bi2223 shows the presence of Bi-O {\it pockets} in the FS due to the interaction between Cu-O and Bi-O4 planes by the mediation of O3, in total disagreement with the experimental reports~\cite{5a}. In the second, the Pb substitution (20-25 \%) on the Bi sites in Bi2223 produces a significant reduction in the interaction between Cu-O and Bi-O4 planes but still the contribution of the Bi/Pb-O states at $E_F$ around point $\overline {\text {M}}$ remains. The third one, containing 26\% Pb or more, shows an almost null interaction between the Cu-O and Bi-O4 planes and, consequently, the contribution of the Bi-O states at $E_F$ is minimal and the Bi-O {\it pockets} are not observed on the FS. The band structures of Pb-free and Pb-doped Bi2223 with Pb concentration of $x=$ 0.0 (Undoped), 0.25 (25\% Pb) and 0.26 (26\% Pb) are shown in Fig.~\ref{B}.

For Pb-free Bi2223, see Fig.~\ref{B}(a), partially filled Bi-O bands around the anti-nodal point $\overline{\text{M}}$ were observed which hybridize with the Cu-O bands. Notice the metallic character of the Bi-O plane and the presence of Bi-O {\it pockets} in the FS, in disagreement with experimental reports~\cite{5a}. A detailed analysis of this band structure can be found in the Ref.~\cite{3a}. The Fig.~\ref{B}(b) shows the band structure of Bi2223 when the 20\% of Bi is replaced by Pb ($x=$ 0.20). A rigid displacement of the Bi/Pb-O bands toward higher energies is observed. These bands are partially filled and contribute at $E_F$. The hybridization with the Cu-O bands is apparently null. The extended van Hove singularities (VHSs) appear at binding energies of -0.35 and -0.15 eV.

When Pb concentration is equal to or greater than 26\% ($x=$ 0.26), see Fig.~\ref{B}(c), the empty Bi/Pb-O bands are lifted rigidly above $E_F$ with no contribution at $E_F$ around the point $\overline{\text{M}}$ solving apparently the Bi-O {\it pockets} problem. The Cu-O bands do not suffer any significant effect due to Pb doping keeping the reported characteristic behaviour of cuprates. An identical result (Fig.~\ref{B}(c)) was reported in a previous paper~\cite{3ax} where without Pb doping with a simple displacement of the O3 atom instead towards the Bi one (away from Cu2), the same movement of the Bi-O bands was induced avoiding completely its contribution at $E_F$. This atomic displacement avoided the charge transfer between Cu2 and Bi through O3, eliminating the metallic character of Bi-O4 plane. As mentioned before,the presence of the Bi-O {\it pockets} are associated to the interaction between Cu-O and Bi-O planes. The presence of Pb in Bi2223 using the VCA, can be interpreted as a reduction of electric charge in the Bi that avoids the presence of possible electronic states that participate in Cu2-O3-Bi-interaction, which filled the Bi-O band.

\begin{figure}[!ht]
\begin{center}
\includegraphics[width=1\textwidth]{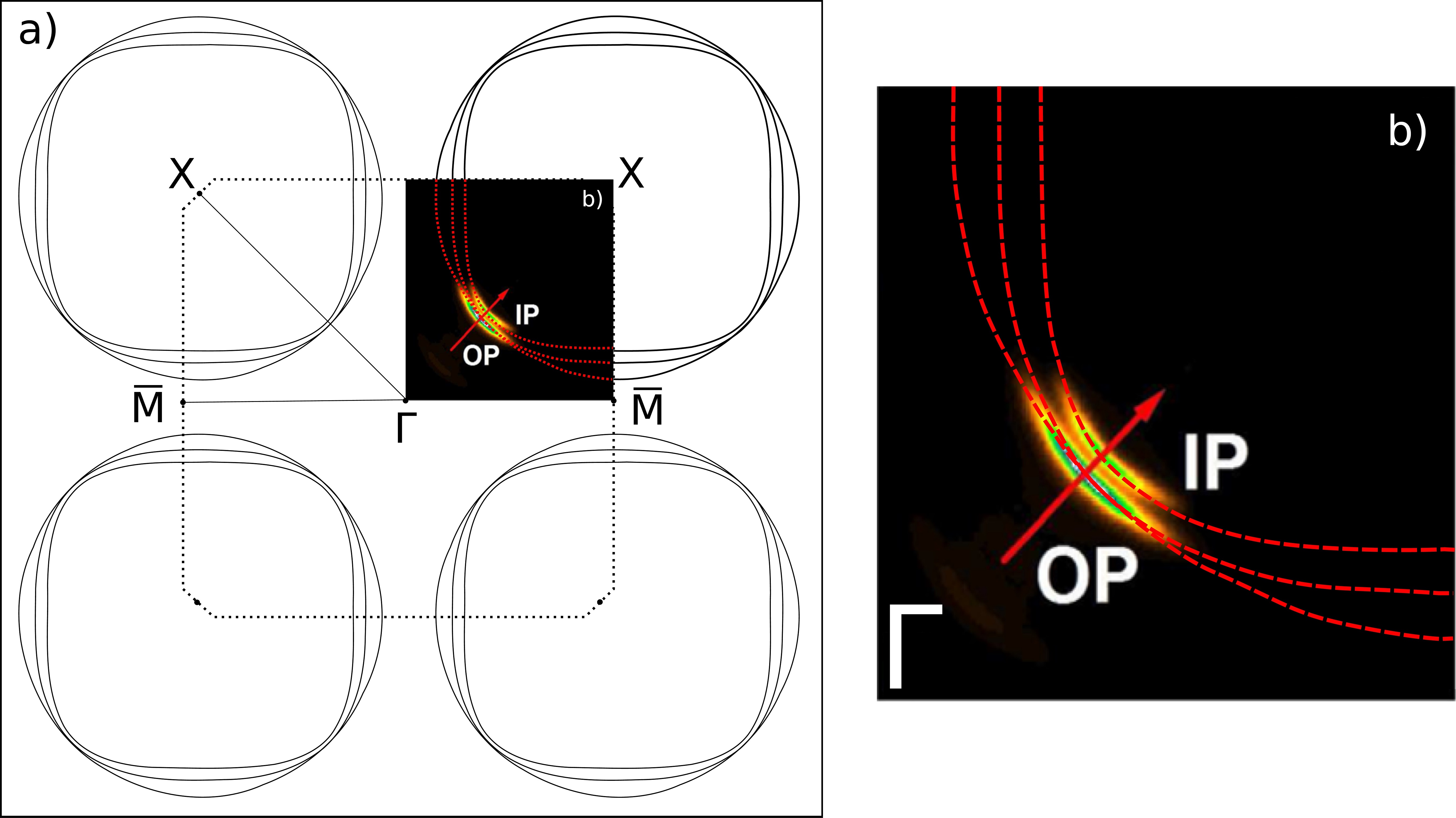}
\caption{(Color online) a) The Fermi Surface (FS) at $k_z=$ 0 of Bi$_2$Sr$_2$Ca$_2$Cu$_3$O$_{10}$ compound in an extended zone scheme, for Pb concentration of 26\% ($x=$ 0.26). b) The experimental FS reported by Ideta {\it et al.} measured in the nodal direction by angle-resolved photoemission spectroscopy (ARPES)~\cite{5a}, compared with our results (red dashed line).}
\label{FS} 
\end{center}
\end{figure}

Fig.~\ref{FS} shows the calculated FS for Pb-doped Bi2223 at 26\% compared to the experimental FS reported by Ideta {\it et al.} which was measured in the nodal direction by angle-resolved photoemission spectroscopy (ARPES)~\cite{5a}. This experiment revealed the presence of two surfaces which were assigned to the outer copper Cu2-O2 planes (OP) and to the inner copper Cu1-O1 plane (IP).

The calculated FS (see Fig.~\ref{FS}(a)) shows three concentric closed surfaces around point X of the IBZ, without the presence of so-called Bi-O {\it pockets} around the $\overline{\text{M}}$ point. Two of this surfaces show quasi-degeneration in the nodal direction which are associated with Cu2-O2 (OP) planes, while the other surface come from the Cu1-O1 plane (IP). These results show a very good agreement with experiment~\cite{5a} (see Fig.~\ref{FS}), as well as with studies of nuclear magnetic resonance (NMR) where it was reported that the hole concentration of OP is larger than IP~\cite{25x,26x}.

\begin{figure}[!ht]
\begin{center}
\includegraphics[width=1\textwidth]{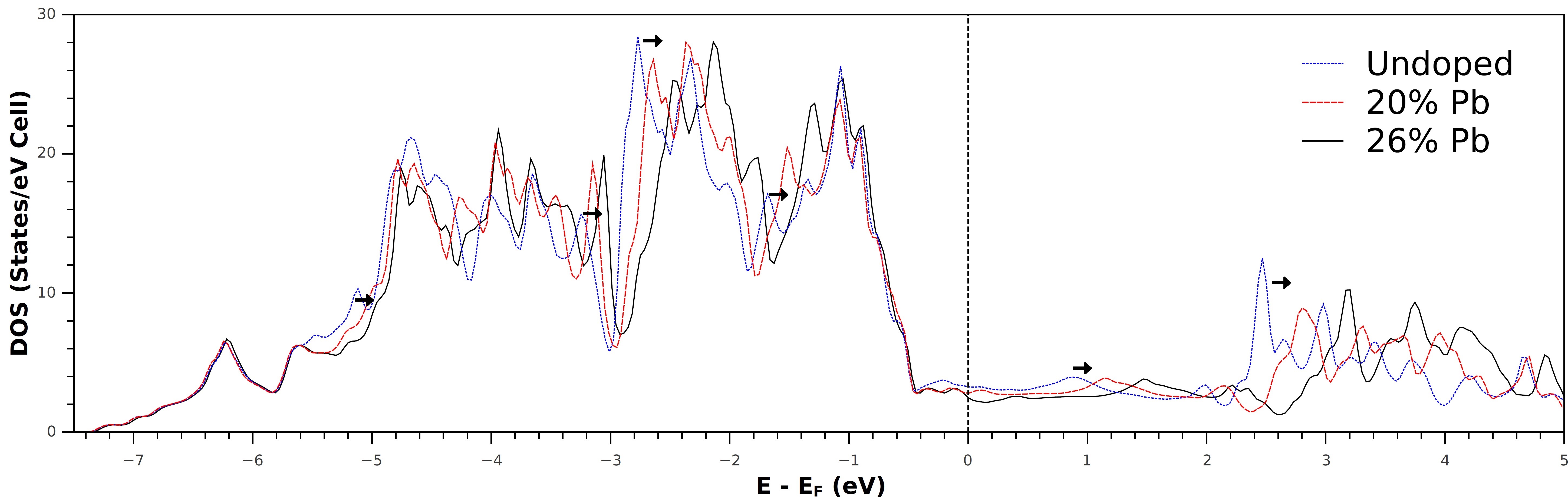}
\caption{(Color online) Pb doping effects on the total density of states (DOS) in the Bi$_2$Sr$_2$Ca$_2$Cu$_3$O$_{10}$ compound, for Pb concentrations of a) 0.0 (Undoped), b) 0.2 (20\% Pb) and c) 0.26 (26\% Pb). }
\label{dos1} 
\end{center}
\end{figure}

The Pb doping effects on the total density of states (DOS) in Bi2223 are presented in the Fig.~\ref{dos1}. These DOS show that the
$E_F$ falls in a region of low DOS, a typical behaviour of these Cu-O based superconductors. As a general behaviour, Pb doping induces a displacement towards higher energies in the DOS, which leads to changes in the contribution of the electronic states at $E_F$. 

\begin{figure}[!ht]
\begin{center}
\includegraphics[width=0.5\textwidth]{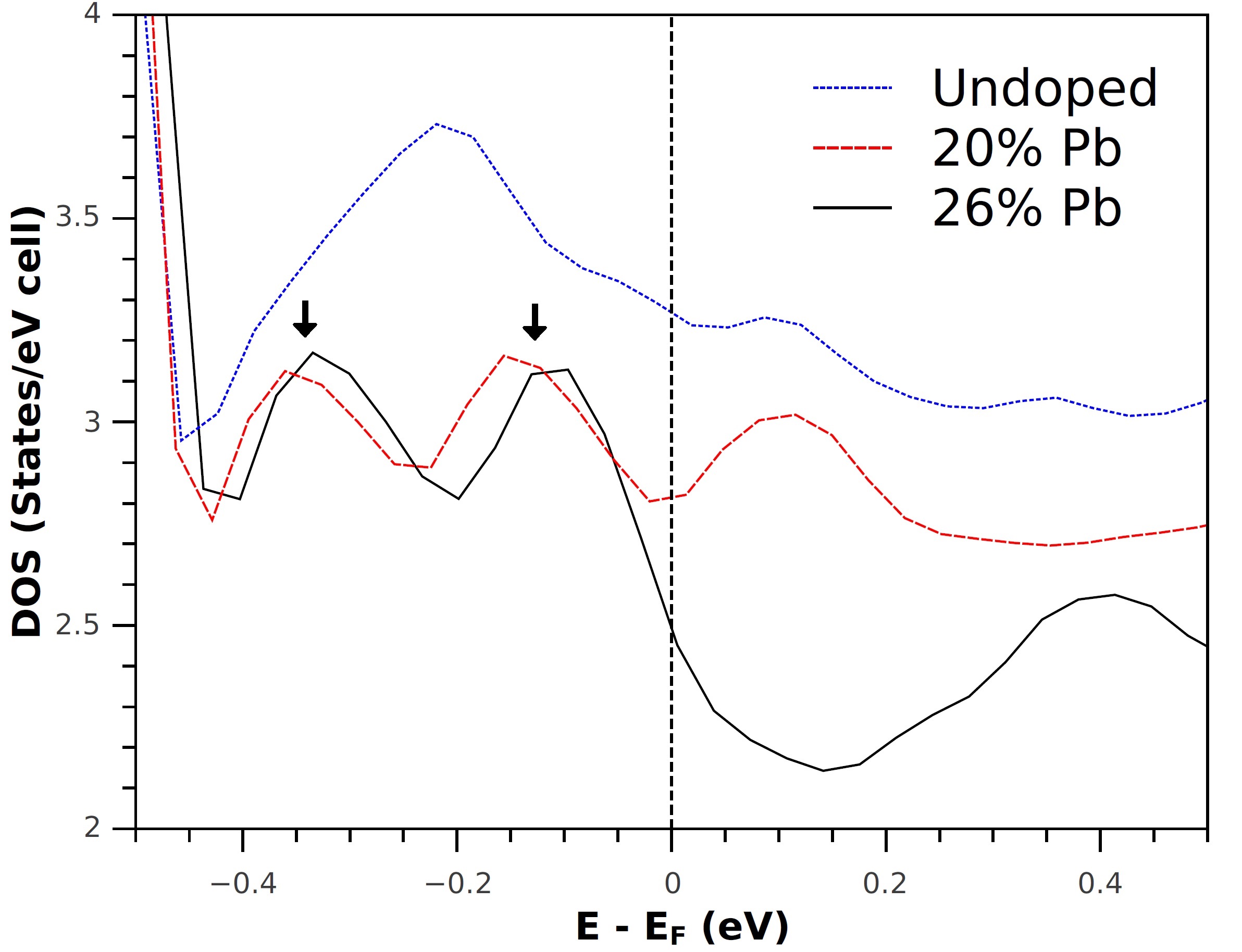}
\caption{(Color online) Pb doping effects on the total density of states (DOS) around $E_F$ in the Bi$_2$Sr$_2$Ca$_2$Cu$_3$O$_{10}$ compound, for Pb concentrations of a) 0.0 (Undoped), b) 0.2 (20\% Pb) and c) 0.26 (26\% Pb). Arrows indicate two peaks at binding energies of -0.35 and -0.15 which are associated with the extended van Hove singularities.}
\label{dos2} 
\end{center}
\end{figure}

The Fig.~\ref{dos2} shows that the DOS at $E_F$, N($E_F$), decreases as Pb doping increases in Bi2223. The extended van Hove singularities (VHSs) appear at binding energies of -0.35 and -0.15 eV, respectively, which play an important role in the physics of high-temperature superconductors~\cite{vhs}.

The N($E_F$) in the Pb-free Bi2223 is 3.27 states/(eV cell) which has contributions from Cu-O and Bi-O planes, mainly of Cu $d_{x^2-y^2}$-O $p_{x,y}$ y Bi $p_{x,y}$-O3 $p_{x,y}$-O4 $p_{x,y}$ states, showing the metallic character of Bi-O plane. The contribution of the Bi-O states at $E_F$ (Bi-O {\it pockets}) is due to the interaction between Cu-O and Bi-O planes, as shown in the hybridization of their respective energy bands, see Figure~\ref{B}(a). This N($E_F$) also has a small contribution of Cu $d_{z^2}$-O $p_{z}$ and Bi $p_{z}$-O3 $p_{z}$-O4 $p_{z}$ states.

At 20\% Pb, the contribution of the Cu-O and Bi-O planes at $E_F$ decreases to 7\% and 56\% respectively, i.e., the interaction between Cu-O and Bi-O states decreases significantly. The Bi $p_{x,y}$-O3 $p_{x,y}$-O4 $p_{x,y}$ states are the ones that decrease the most their contribution at $E_F$ which becomes 2.81 states/(eV cell) meanwhile Bi $p_{z}$-O3 $p_z$-O4 $p_z$ states are not affected by the presence of Pb in the structure. With increasing Pb doping up to 25\% the electronic properties show no significant changes.

Further, the N($E_F$) is 2.49 states/(eV cell) when Pb doping is 26\%; the contribution at $E_F$ of Bi $p_{x,y}$-O3 $p_{x,y}$-O4 $p_{x,y}$ states is almost null, although a contribution of Bi $p_z$-O3 $p_z$-O4$ p_z$ states is observed, representing a little more than 1\% of the total contribution at $E_F$, which can be interpreted as a slight metallic character of Bi-O plane. For 26\% Pb doping, this metallic character is maintained. The presence of these Bi-O states comes from the energy bands that intersect at $E_F$ in the $\overline{\text{M}}$-X direction, see Fig.~\ref{B}(c).

\begin{figure}[!ht]
\begin{center}
\includegraphics[width=0.75\textwidth]{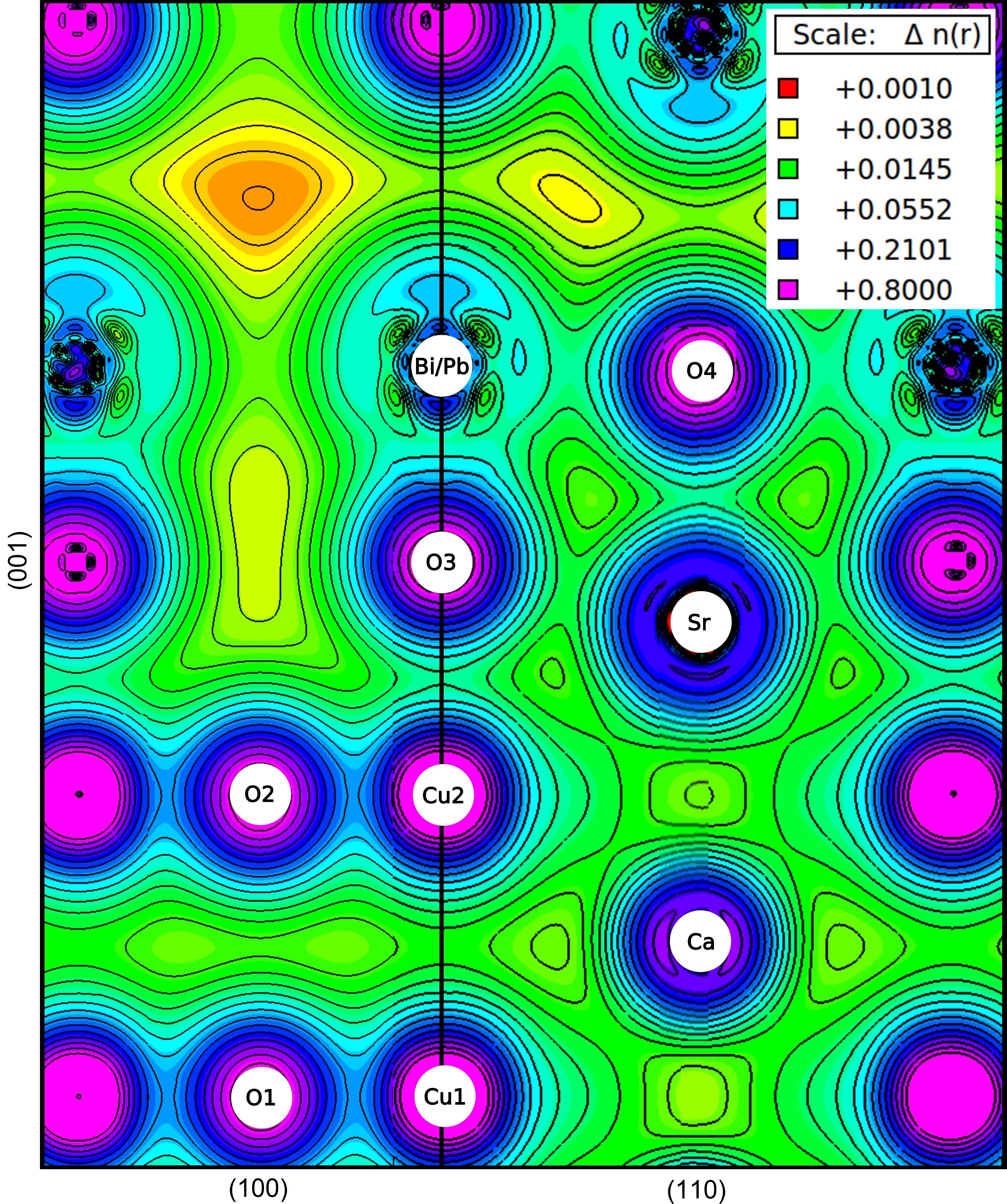}
\caption{(Color online) Contours of constant values of the logarithm of the valence density of 26\% Pb-doped Bi2223, in two high-symmetry planes as labeled, obtained with XCrySDen program~\cite{xd}. Scale units of $e/a.u.^2$.} 
\label{de3} 
\end{center}
\end{figure}

The charge density calculated on (100) and (110) high-symmetry planes in the 26\% Pb-doped Bi2223 presented in Fig.~\ref{de3} reflects the strong ionic nature of Bi/Pb atoms, as the ionic character of Sr and Ca. 

The charge density between Cu1(2) and O1(2) reveals a strong metallic character, typical of this cuprates. There are regions of small charge density in the order of $1.6\times^{-2}$ to 3.3 $\times^{-2}$ $e/au^2$, between (Bi/Pb)-O3-Cu2 and (Bi/Pb)-O4 bonds representing a slight metallic character in these directions.

Experimental results using scanning tunneling microscopy (STM), show that the Bi-O planes are non-metallic in Bi2212~\cite{s1,s2,s3}. To the best of our knowledge there are no experiments that define the metallic or non-metallic character of the Bi-O planes in Bi-2223, except for the report presented by K. Asokan et al. that using X-ray absorption near edge structure (XANES), show a metallic character of Bi-O planes for Bi-2223 and Bi-2212 ~\cite{s4}. t\emph{T}he last one is in disagreement with the previous works just mentioned. However, the nature of Bi-O planes character for Bi-2223 needs to be tested with more experiments.

We reproduce the calculations reported by H. Lin {\it et al.}~\cite{8a} and effectively it was found that for Pb doping concentration equal to or more than 22\% the Bi-O bands move toward higher energies avoiding its contribution at $E_F$ on the point $\overline{\text{M}}$ (no Bi-O {\it pockets} in FS). Although it was also found in this case that, as in Bi2223, a slightly metallic character in the Bi-O bonds due to contribution of Bi-O states at $E_F$ on $\overline{\text{M}}$-X direction is observed, in disagreement with the Bi2212 experimental reports. Thus, Pb doping in Bi2212 and Bi2223 structures certainly avoids the presence of the {\it pockets} at  FS, although the metallic character of the Bi-o planes remains.

\section{Conclusions}

In this paper, we presented a study of Pb doping effects on the structural and electronic properties of (Bi$_{1-x}$Pb$_x$)$_2$Sr$_2$Ca$_2$Cu$_3$O$_{10}$ for $ 0 \leq x \geq 0.35$, using the Local Density (LDA) and Virtual Crystal (VCA) approximations within the framework of the Density Functional Theory (DFT) with the Wien2k code, taking as reference the method applied in Bi2212 by H.Lin {\it et al.}~\cite{7a,8a} to study of Bi2223. 

Our results show significant changes in Bi2223 structural properties when Pb is incorporated as a dopant. These changes are associated to the presence or absence of the Bi-O {\it pockets} at the Fermi surface (FS). The most relevant feature is observed in Cu2-O3 distance, which becomes lower when the Bi-O {\it pockets} disappear from the FS. Although differences between experimental parameters, $a$ and $c$, and the calculated ones are not greater than 1\%, appreciable discrepancies in the interatomic distances are observed, as in the case of the Cu2-O3 distance which is 0.6 to 0.8 \AA ~greater than the experimental values reported.

It was observed, as a general that Pb doping in Bi2223 lead a rigid displacement of the Bi/Pb-O bands toward higher energies with a null contribution at the Fermi level in the high symmetry point $\overline{\text{M}}$ of IBZ for Pb doping concentration equal to or more than 26\%. This displacement, avoids the presence of the so-called Bi-O {\it pockets} at the FS, in good agreement with angle-resolved photoemission spectroscopy (ARPES) and nuclear magnetic resonance (NMR) experiments. A slightly metallic character of the Bi-O planes still remains due to the contribution of the Bi $p_z$ -O3 $p_z$ -O4$ p_z$ states at $E_F$ in the $\overline{\text{M}}$-X direction. The Cu-O bands do not suffer any significant effect due to Pb doping, keeping the reported characteristic behaviour of the cuprates.

The metallic character of the Bi-O bonds is also observed in Bi2212. this result disagree with some experimental reports. Thus, the absence of the Bi-O pockets does not guarantee the non-metallic character of the Bi-O planes in either Bi2212 or Bi2223 compounds.

\end{document}